\documentclass{iopart} 
\usepackage{iopams} 
\usepackage{xspace}
\usepackage{url}

\usepackage{graphicx}

\usepackage{color}

\newcommand{\w}{\ensuremath{\omega}}
\newcommand{\I}{\ensuremath{{\rm i}\,}}
\newcommand{\zr}{\ensuremath{z_{\rm R}}}

\newcommand{\HG}{Hermite-Gauss\xspace}

\let \IG \includegraphics

\begin{document}

\title[Advanced Virgo arm cavity design] 
{Using the etalon effect for in-situ balancing of  the Advanced Virgo arm cavities}

\author{S.~Hild$^1$, A.~Freise$^1$, M.~Mantovani$^3$,
 S.~Chelkowski$^1$, J.~Degallaix $^2$ and R.~Schilling$^2$} 
\address{$^1$ School of Physics and
Astronomy, University of Birmingham, Edgbaston, Birmingham B15 2TT, United Kingdom}
\address{$^2$ Max-Planck-Institut f\"ur Gravitationsphysik
(Albert-Einstein-Institut) and Leibniz Universit\"at Hannover,
Callinstr. 38, D--30167 Hannover, Germany.}
\address{$^3$ European Gravitational Observatory (EGO), Via E Amaldi, I-56021 Cascina (PI), Italy}

\ead{hild@star.sr.bham.ac.uk}

\begin{abstract}

 Several large-scale interferometric gravitational-wave detectors use resonant arm cavities to
 enhance the light power in the interferometer arms. These cavities are based on different
 optical designs: One design uses wedged input mirrors to create additional optical pick-off
 ports for deriving control signals. The second design employs input mirrors without wedge and
 thus offers the possibility to use the etalon effect inside the input mirrors for tuning
 the finesse of the arm cavities. In this article we introduce a concept of maximized
 flexibility that combines both of these options, by featuring wedges at the input mirrors and
 using the etalon effect instead in the end mirrors. We present a design for the arm cavities of
 Advanced Virgo.  We have used numerical simulations to derive requirements for the
 manufacturing accuracy of an end mirror etalon for Advanced Virgo.  Furthermore, we give
 analytical approximations for the achievable tuning range of the etalon in dependence on the
 reflectance, the curvature and the orientation of the etalon back surface.

\end{abstract}



\section{Introduction}
The first generation of large-scale interferometric gravitational-wave detectors \cite{geo, virgo,
 tama, ligo} has successfully been constructed. Simultaneous long duration data-takings have
 been performed. Currently most of the detectors undergo minor upgrades including only a few
 hardware changes. However, in a few years major upgrades will be performed leading to the
 so called second generation gravitational-wave detectors. The construction of these
 instruments, like Advanced Virgo \cite{advvirgo} and Advanced LIGO \cite{advligo}, will include
 major changes of almost all subsystems. In  Advanced Virgo, amongst others, the beam
 geometry inside the arm cavities will be changed and new mirrors will be installed. This
 provides us with the possibility to introduce and evaluate new concepts for the optical layout of
 the arm cavities. As we will show it is possible to optimize the arm cavity design by using
 wedged input mirrors and employing the etalon effect at the end mirrors in order to realize arm
 cavities with tunable finesse, i.e. adjustable losses.

We start in Section \ref{sec:motivation} by giving a brief motivation for using arm cavties providing in-situ balancing.
 In Section \ref{sec:principle} we give a brief description of the advantages and disadvantages
 of using either wedged or unwedged input mirrors, before we propose a concept which
 combines the advantages of both options. Analytical calculations and numerical simulations
 are used  in Section \ref{sec:application}  to show how the tunable range is reduced by
 imperfections of the etalon, such as curvature mismatch or nonparallelism of the two etalon 
surfaces. In Section \ref{sec:curv_and_para} we analyse the influence of imperfect optics onto
 other  subsystems of Advanced Virgo, like for instance the auto-alignment system. A summary
 is given in Section \ref{sec:summary}.   

\section{Motivation for arm cavities providing in-situ balancing}
\label{sec:motivation}

A Michelson interferometer with unbalanced arm cavities  (arm cavities with differing finesse 
and/or losses) is  undesirable for the following reasons: 
\begin{itemize}
\item Due to the different light storage times  inside two arm cavities, the common mode
 rejection for several technical noise sources is reduced resulting in a stronger noise coupling
 into the gravitational wave channel \cite{Somiya06, Somiya07b}. 
\item The dark fringe contrast is reduced resulting in an increased waste light level at the
detection port and a reduced power recycling gain.
\item The coupling of radiation pressure noise originating from laser intensity noise into
 the main interferometer output is increased.  
\end{itemize}
 
In practice, mismatches of the reflectivities, the cleanliness and the surface roughness of the
 mirrors limit the achievable balance of the two arm cavities.
Since the mirrors of future gravitational wave detectors are likely to be suspended by quasi
-monolithic suspension (instead of metal wire loops) it will be risky to properly clean them after 
installation. Also exchanging a mirror will take much longer than in first generation instruments 
and would therefore cause increased downtime of the instrument.

Hence, it might be beneficial to utilise the etalon effect in one or both of the cavity mirrors in
 order to balance the two arm cavities in-situ. Balancing the interferometer arms might
 significantly reduce the technical problems listed above and therefore speed up the
 commissioning of the second generation instruments.
svn

\section{Principle optical layout of the Advanced Virgo arm cavities}
\label{sec:principle}

\begin{figure}[htb]
\begin{center}
\IG [scale=0.75] {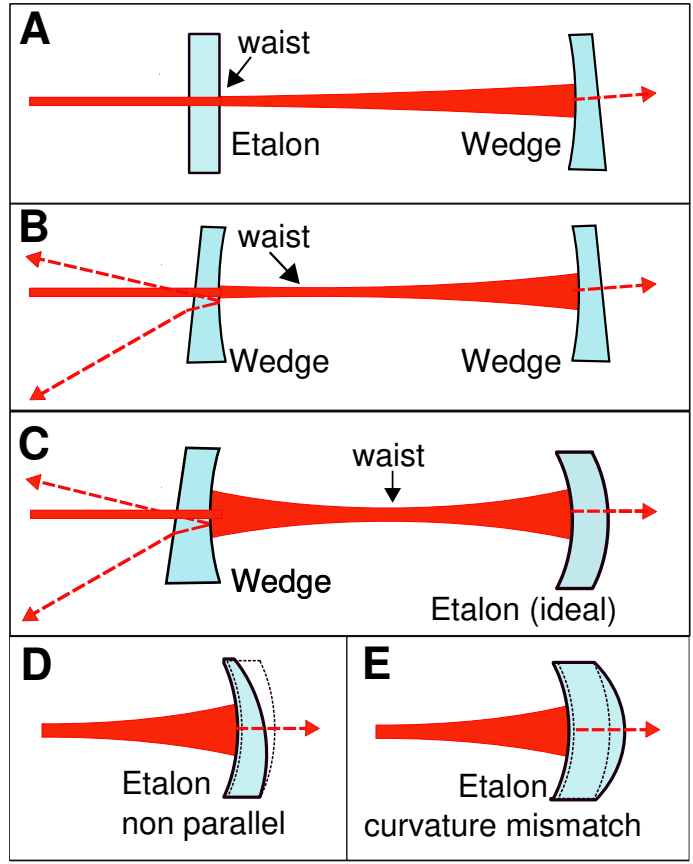} \\
\end{center}
\caption
{The plot shows various arm cavity configurations used by currently operating interferometers, as well as potential configurations for future detectors. From the left side the light enters the arm cavity formed by the input mirror (left) and end mirror (right). A) Initial Virgo configuration: The input mirror is a flat/flat etalon, while the end mirror is curved. B) Initial LIGO configuration: In contrast to A) the input mirror features a wedge (and is slightly curved) . C) Design of the  Advanced Virgo arm cavities as proposed by the authors: Both mirrors are identically curved in order to shift the beam waist to the centre of the cavity. The input mirror features a wedge, while the etalon effect of the end mirror is used. In order to maximize the etalon effect both surfaces of the end mirror have to be curved identically. D) and E) show a non perfect etalon. The influence of both illustrated imperfections, non parallelism (D) and curvature mismatch (E) will be analysed in Section \ref{sec:curv_and_para} of this paper.}
\label{fig:layouts}
\end{figure}

\subsection{Etalon to tune the mirror reflectance}
The arm cavity configuration of initial Virgo is shown in Figure \ref{fig:layouts}\,A. The end mirror is a wedged curved/flat mirror with an anti reflective (AR) coating at the back surface. The input mirror is a plane/plane mirror (featuring no wedge) with an AR coating and the mirror coating on the left and right surface, respectively. Due to the etalon effect the  overall reflectance of the input mirror changes depending on the microscopic tuning of the etalon, i.e. the optical path length inside the etalon substrate.

\subsection{Wedges and pick-off beams}

For the arm cavities of initial LIGO a different approach was chosen. As indicated by Figure
 \ref{fig:layouts}\,B the input mirror is wedged, thus any potential etalon effect is suppressed, so
 that even without any temperature control of the input mirror the cavity finesse is stable. This
 concept is simple and robust, but  lacks any possibility to balance the arm cavities.

 The main benefit of featuring wedged input mirrors is the availability of additional pick off beams from the input mirror AR coating.  These beams can be used for deriving further control signals. For advanced gravitational-wave detectors the availability of many control ports with different signal content seems highly desirable, especially because all second generation instruments will employ Signal-Recycling \cite{Meers88}, which  strongly increases the
 complexity of the length sensing and control systems \cite{hrg_Dual}, \cite{40m}.   

\subsection{Proposed optical layout for arm cavities in advanced interferometers}

While the optical design for Advanced Virgo is not yet complete, it has been decided to to shift
 the beam waist to the center of the cavity. This will maximise the beam spot size on both mirrors
 thus reducing the effects of thermal noise from the mirror substrates and the dielectric coatings.
We can summarize  the requirements and boundary conditions for an optimal optical design 
of the arm cavities for the Advanced Virgo advanced detector as follows:

\begin{itemize}
\item Wedged input mirrors to create additional pick-off beams
\item Tunable cavity finesse (tuning range of approximately a few percent)
\item Identical radius of curvature of input and end mirror (Beam waist centered in the arm cavity)
\end{itemize}

Figure \ref{fig:layouts}\,C shows a schematic of the arm cavity concept that we propose. The 
 input mirror is flat/curved featuring a wedge, thus two pick-off beams are created. Since no
 etalon effect is present, the overall reflectance of the input mirror is independent of the
 temperature inside the substrate. Having no etalon effect in the input mirror has the advantage
 that the requirements for the performance of a thermal compensation system \cite{lawrence}
 can be relaxed, i.e. only thermal lensing needs to be compensated for. 
The end mirror has no wedge\footnote{At the end mirror no additional pick off beams are required because the transmitted beam is already accessible.} and serves as etalon. In order to provide a reasonable tuning range of the overall reflectance of the end mirror, its back surface carries a dielectric coating of moderate reflectance (tens of \%). The best etalon performance can be achieved by also curving the back surface of the etalon, ensuring perfect mode matching of the cavity beam into the etalon. 

The performance of such an ideal etalon configuration is easily computable and in principle no additional problems should occur, assuming the optical path length, i.e. the temperature inside the etalon substrate is well controlled (see Section \ref{sec:op_phase}).
However, in reality we have to deal, at least to some extent, with a non perfect etalon.  Etalon
imperfections originating from limited accuracy in the manufacturing process of the etalon, lead
 to a non perfect wavefront matching inside the etalon. In addition, light absorption at the front
 surface coating can induce thermal deformation and thermal lensing inside the etalon. Any
 imperfections of the etalon do not only reduce the available tuning range of the etalon
 reflectance, but also influence the light field inside the arm cavity. An analysis of potential
 distortion effects originating from a non perfect etalon is carried out in the Section
 \ref{sec:curv_and_para}.

\section{Application of an ideal etalon in Advanced Virgo}
\label{sec:application}
 In this Section we apply the concept proposed in the previous section to the design of the Advanced Virgo detector. We use analytical calculations to describe the tuning range of the etalon in dependence of the reflectance of the etalon back surface. In addition we analyse optical phase noise originating from temperature fluctuations inside the etalon.
The optical parameters of Advanced Virgo, used throughout this paper, are displayed in Table
 \ref{tab:adv_para}. Please note that we give transmittance, $T_i$, and loss, $L_i$, in power,
 while the reflectance, $\rho_i=\sqrt{1-T_i-L_i}$, is given in amplitude.

\begin{table}\label{tab:adv_para}
\begin{center}
\begin{tabular}{|rl|l|}
\hline
input mirror &  $T_1$ & 0.007\\
& $L_1$ & 50\,{\rm ppm} \\
 & $R_{C1}$ & 1910\,m\\
\hline
end mirror front &  $T_2$ & 50\,{\rm ppm}\\
& $L_2$ & 50\,{\rm ppm} \\
 & $R_{C2}$ & 1910\,m\\
\hline
end mirror back &   $T_3$ & variable (0.8 -- 0.98)\\
&$L_3$ & {\rm none} \\
 & $R_{C3}$ & 1910\,m (ideal)\\
\hline
 cavity length & &3000\,m \\
\hline
 mirror diameter && 0.35\,m \\
\hline
\end{tabular}
\caption{Set of Advanced Virgo arm cavity parameter we use in this article. Loss, $L_i$ and transmittance, $T_i$,  are given in power, 
the respective amplitude reflectivities calculates as
$\rho_i=\sqrt{1-T_i-L_i}$. The radii of curvature of the mirror surfaces are indicated by $R_{Ci}$. (In the analytical description we assume
mirrors of infinite size, while the FFT simulations uses mirrors of
35\,cm in diameter.)}

\end{center}
\end{table}

\subsection{Achievable tuning range of an ideal etalon}

\begin{figure}[htb]
\begin{center}
\IG [scale=0.36] {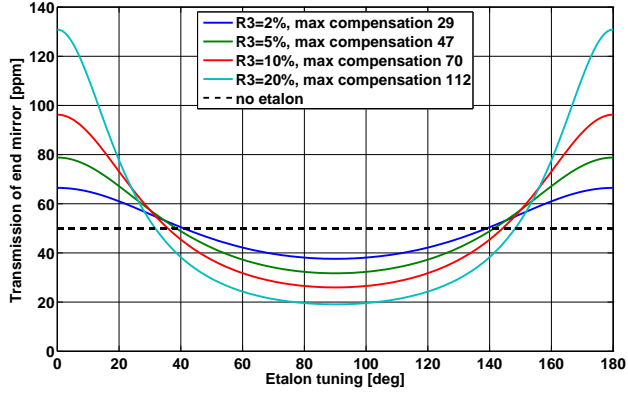} \\
\end{center}
\caption[amplitude reflectance of the etalon]
{The plot shows the power transmittance of the end mirror using the etalon effect as a function 
of the tuning angle $\phi$. The etalon is assumed to have a power reflectance of
 $R_2=1-50$\,ppm
at the front face and of $R_3$ at the back surface.}
\label{fig:ETR1}
\end{figure}

The reflectance of a lossless etalon, $\rho_E$ can be written as:
\begin{equation}
\rho_E=\rho_2-\frac{\rho_3 (1-\rho_2^2)}{\exp(-2\I\phi)-\rho_2\rho_3}
\end{equation}
with $\rho_2$ and $\rho_3$ being the amplitude reflectance of the etalon front and back
surfaces, respectively, and $\phi$ the tuning of the etalon with $\phi=180$\,deg 
referring to one free spectral range.
The absolute value of $\rho_E$ 
can be used to determine the power reflectance or transmittance of the etalon.
Figure~\ref{fig:ETR1} shows the overall transmittance of a cavity end mirror etalon using the
 parameters given in Table \ref{tab:adv_para}.
The difference between the maximum and minimum transmittance can be called `maximum
 compensation' and describes what amount of differential 
losses in two arm cavities can compensated for by tuning the end mirror etalon.

\begin{figure}[htb]
\begin{center}
\IG [scale=0.4] {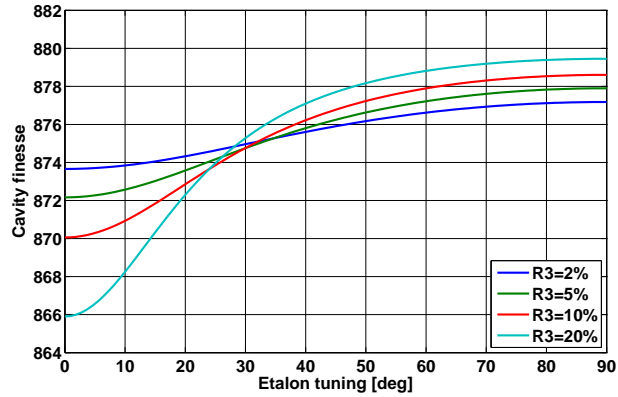} \\
\end{center}
\caption[arm cavity finesse]{Arm cavity finesse in dependence of the etalon tuning and the
 power reflectance $R_3$ of the etalon back surface.}
\label{fig:PC1}
\end{figure}

The Finesse, $F$, of  the arm cavity in dependence of the etalon tuning  and the reflectance of
 the etalon back surface can be computed using:  
\begin{equation}\label{eq:pcirc}
F=\frac{\pi}{2}\arcsin^{-1}\left(\frac{1-\rho_1 \rho_E}{2\sqrt{\rho_1 \rho_E}}\right)
\end{equation}
Figure~\ref{fig:PC1} shows the Finesse, $F$, of the cavity as 
a function of the etalon tuning using the mirror parameters shown
in Table~\ref{tab:adv_para}.

\subsection{Optical phase noise}
\label{sec:op_phase}

\begin{figure}[htb]
\begin{center}
\IG [scale=0.4] {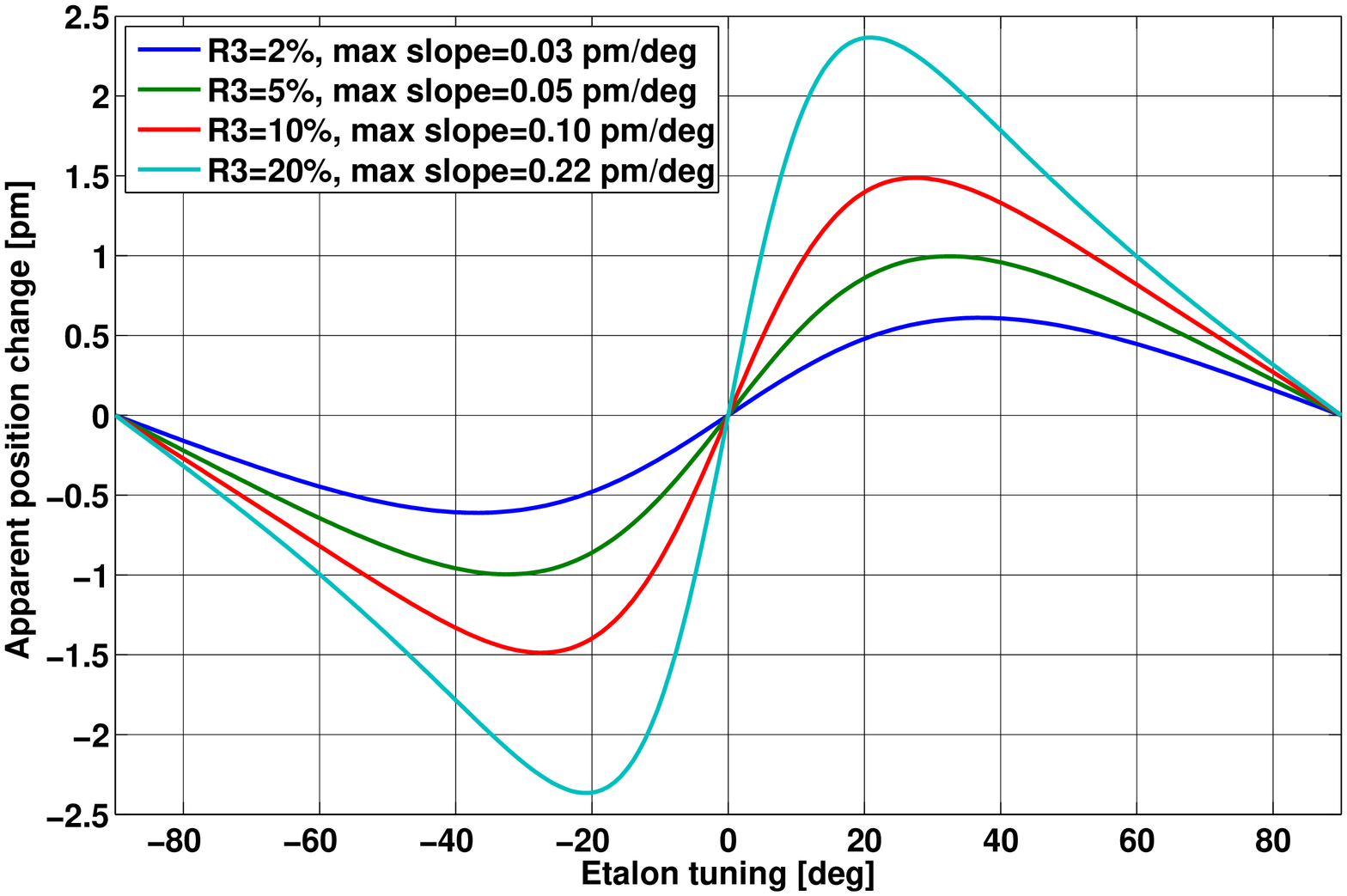} \\
\end{center}
\caption[apparent position change]
{The plot shows the apparent position change of the end mirror as a function of the
etalon tuning. The maximum slope around the zero tuning point is reported in the label.}
\label{fig:ETR2}
\end{figure}

The overall reflectance of the etalon differs from that of an ordinary mirror not only in amplitude
but also in phase. This can be interpreted such that the apparent position of the
end mirror is influenced by the etalon tuning. This of course couples etalon
fluctuations into optical phase noise, i.e. apparent displacement noise of the test masses
 at the end of the arm cavities. The apparent position change of the mirror
can be computed as:
\begin{equation}
\Delta x_E=\frac{\lambda}{4\pi}\arctan\left(\frac{\Im(\rho_E)}{\Re(\rho_E)}\right)
\end{equation}
Figure~\ref{fig:ETR2} shows the apparent position change as a function of the etalon tuning
for a few examples. Considering a `worst case scenario' of an etalon back surface coating 
of a power reflectance of
 $R_3 = 0.2$ and an etalon operating point with the strongest slope of the optical phase noise,
 $n_p = 0.22\,$pm/deg, we can calculate the corresponding displacement noise spectrum,
 $\tilde{x}_E$, in dependence of the amplitude spectral density of the effective temperature
 fluctuations, $\tilde{T}_E(f)$, inside the etalon:

\begin{equation}
\tilde{x}_E(f) = \tilde{T}_E(f) \cdot \frac{dn}{dT} \cdot l_{E} \cdot n_p ,
\end{equation}
where $l_E$ is the thickness of the etalon and $dn/dT$ is the change of the index of refraction
of the etalon substrate with temperature.
As shown in \cite{Virgo-note-phase-noise}  the effective temperature
 fluctuations of the etalon have to be kept below $10^{-10}\,\textrm{K}/\sqrt{\rm Hz}$ for most
 frequencies within the detection band in order to avoid spoiling the Advanced Virgo sensitivity.
 We can compare this requirement to the actually present temperature fluctuations, assuming 
that these are (for frequencies in the detection band) entirely driven by Brownian fluctuations.
Using Equation 5.2 from \cite{Braginski04} we can describe the effective temperature
 fluctuations of the etalon to be:
\begin{equation}
\tilde{T}_{\rm E}(f) = \sqrt{ \frac{4 k_b\, T^2 \kappa}{(\rho\, C)^2  l_{\rm eta}} \frac{1}{\pi R^4_b (2\pi f)^2}},
\end{equation}
where $k_b$ is the Boltzmann constant, $T$ the temperature of the Etalon, $\kappa$ the
 thermal conductivity, $\rho$ the density of the etalon, $C$ the heat capacity and $R_b$ the
 radius of the laser beam inside the etalon substrate.
 It turns out that the Brownian temperature fluctuations are at least four orders of magnitude
 below the temperature requirement of the etalon \cite{Virgo-note-phase-noise}.

\section{Analysis of etalon imperfections} 
\label{sec:curv_and_para}

So far we assumed a geometrically perfect etalon with parallel
surfaces. In practise the etalon substrates will deviate from a perfect shape.
The geometrical deviations with the largest impact on the reflected light are
\begin{itemize}
\item deviation of parallelism (i.e. relative misalignment)
\item mismatch in spherical curvature
\end{itemize}
The first case is illustrated in Figure \ref{fig:layouts}\,D. The two etalon 
surfaces have identical
curvature, but are misaligned in respect to each other, i.e. not parallel. 
Figure \ref{fig:layouts}\,E shows the case of curvature mismatch of the two etalon surfaces.

In order to evaluate the effects of potential etalon imperfections on the arm cavity 
performance for Advanced Virgo we use two different methods: Method 1 makes use   
of Hermite-Gauss coupling coefficients (HGCC). For Method 2 we use  
FFT-based numerical simulations. In addition, we have used the interferometer simulation
software \textsc{Finesse} \cite{finesse1, finesse2} to cross-check the results of the other 
two methods. The results from \textsc{Finesse} agreed well with the other two methods, 
and are hence omitted from the plots in this section for clarity. 

A short description of each of the two methods is given in the Sections \ref{sec:coup} and
\ref{sec:fft}. We use both methods to calculate how the  achievable tuning range of an 
etalon is influenced by its imperfections (Section \ref{sec:tune_imper}). Finally, using the
example of the alignment sensing and control system, we evaluate how the etalon
imperfections can potentially influence other subsystems of the Advanced Virgo detector 
(Section \ref{sec:align}).


\subsection{Method 1: Hermite-Gauss Coupling Coefficients}
\label{sec:coup}

In this section we compute coupling coefficients which determine how much 
light from an TEM$_{00}$ mode incident on the etalon will be scattered into
higher order modes due to the imperfection of the etalon back surface.
In a first approximation we consider the light such scattered into
higher order modes as additional losses. This is an approximation because
in reality the coupling into higher order modes can be enhanced or reduced
by the arm cavity resonance. However, as we show in the following, analytic
expressions derived with this method match the results from numerical 
simulations very well.

A beam profile of a fundamental Gaussian beam 
(for a beam with a given wavelength $\lambda$) can be
completely determined by two parameters: the size of the minimum spot size
$w_0$ (called \emph{beam waist}) and the position $z$ of the beam waist
along the $z$-axis. A convenient method of expression of these two parameters is the 
 \emph{Gaussian beam parameter}:
\begin{equation}
q(z)=\I\zr +z=q_0+z,
\end{equation}
with the \emph{Rayleigh range} $z_R$ defined as
\begin{equation}
\zr=\frac{\pi w_0^2}{\lambda}.
\end{equation}
In the following we describe the light field in the arm cavity and the etalon
using \HG modes which are an orthonormal set of functions which 
can be used to describe deviations of the beam from its fundamental Gaussian shape.
 The scalar electrical field describing a beam propagating along the
z-axis can then be written as
\begin{equation}
E(x,y,z)=E_0 \sum_{nm} c_{nm} u_{nm}(x,y,z) \exp(\I\w-kz),
\end{equation}
in which the set of \HG modes $u_{nm}$ is defined by the Gaussian beam parameter
$q$ of the beam and the coefficients $c_{nm}$ determine the mode composition
of~$E$.
We assume that the arm cavity and the incoming light are perfectly matched, and the
light is in a zero order mode described by  $q_c$ being the beam parameter of the cavity
eigenmode. 

When a beam is  passing through  or reflected by 
a spherical surface a beam parameter $q_1$ is transformed to $q_2$.
The transformation of the beam parameter can be performed by the 
ABCD matrix-formalism \cite{siegman}. In our case the
beam reaching the end mirror passes first into the substrate,
which causes the beam parameter $q_c$
to be transformed as:
\begin{equation}
q_t=n\frac{q_c}{\frac{n-1}{R_C}q_c+1}
\end{equation}
with $R_{\rm C}$ the radius of curvature of that surface and $n$ the
index of refraction of the substrate.
Then upon reflection from the back surface\footnote{The short propagation
through the mirror substrate can be neglected in this case.} 
we get a second transformation
\begin{equation}
-q_r^*=\frac{q_t}{\frac{-2}{R_C}q_t+1}
\end{equation}
with the $^*$ indicating the complex conjugate.
In general $q_t$ and $q_r$ will be different, i.e. the incoming and reflected beam will
have a different shape. In order to compute how much of the light will remain in the cavity
 eigenmode we have to describe the reflected 00 mode, $u_{00}(q_r)$ in the base system of
 $u_{nm}(q_t)$. Such a base change can be described by coupling coefficients $k_{nmn'm'}$:
\begin{equation}
u_{nm}(q_t)=\sum_{n'm'}k_{nmn'm'} u_{n'm'}(q_r)
\end{equation}
We are interested how much of the reflected light is coupled back into the fundamental
cavity eigenmode.
The respective coupling coefficient $k_{0000}(q_1,q_2)$ 
can be found in the literature; 
for example, in \cite{bayer} 
the coupling coefficients are defined using mode-mismatch 
parameters $K_0$ and $K$ given as:
\begin{equation}
K=\frac12(K_0+\I K_2),
\end{equation}
where $K_0=(z_{\rm R, 1}-z_{\rm R, 2})/z_{\rm R, 2}$ and $K_2=(z_1-z_2)/z_{\rm R, 2}$.
This can be also written as:
\begin{equation}
K=\frac{\I (q_1-q_2)^*}{2 \Im(q_2)}.
\end{equation}
This parameter is a measure of the mode-mismatch between two TEM$_{00}$ shapes
defined by the Gaussian beam parameters $q_1$ and $q_2$.

If we assume perfect alignment of the front and back surface, but allow different radii of
curvature of the two etalon surfaces,  the coupling coefficient of interest is then given as:
\begin{equation}
k_{0000}(q_1=q_r,q_2=q_t)=\frac{\sqrt{1+K_0}}{1+K^*} .
\end{equation}
Then the effect of the etalon can be  computed as before by replacing
$\rho_3$ by $|k_{0000}|\rho_3$.
If on the other hand the surfaces have matched curvatures but the back surface is
misaligned by an angle $\gamma$ the coefficient is given as:
\begin{equation}
k_{0000}(q_r,\gamma)=\exp\left(-\frac{\pi |q_r|^2\sin^2(2\gamma)}{2\lambda\,\Im(q_r)}\right)
\end{equation}

\subsection{Method 2: FFT based simulations}
\label{sec:fft}

The second method we used to evaluate the influence of imperfections onto the etalon
 performance is based on FFT simulations using the program \emph{WaveProp}.
WaveProp is a  tool for simulating the propagation of  electromagnetic waves, based on
 Discrete Fourier Transformation. The program is available at \cite{Waveprop}.  The
 method of FFT propagation is described, e.g., by Siegmann \cite{siegman} and by Vinet
 \cite{fft}.

\subsection{Tuning range of an imperfect etalon}
\label{sec:tune_imper}

\begin{figure}[htb]
\begin{center}
\IG [scale=0.4] {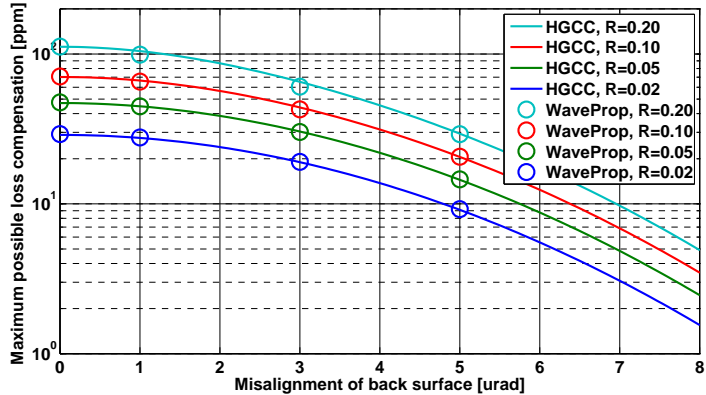} \\
\end{center}
\caption[curvature mismatch]
{The plot shows the maximum loss that can be compensated for by the etalon, as
a function of the etalon rear surface misalignment. The solid traces are computed from
 Hermite-Gauss coupling coefficients (HGCC), while the data points represented by a circle
 are derived from WaveProp simulations.}
\label{fig:ETR4}
\end{figure}

Using the two methods that have been laid out in the previous sections, we can now calculate
how much the etalon performance is reduced depending on its geometrical imperfections.
In general these geometrical imperfections reduce the wavefront overlap inside the etalon and
therefore reduce the achievable tuning range of the etalon reflectivity. 
Figure~\ref{fig:ETR4} shows the maximum loss compensation possible with the parameters 
of an Advanced Virgo arm cavity (see Table \ref{tab:adv_para}). For different reflectances of the
 etalon rear surface we plotted the achievable loss compensation over the misalignment of 
the etalon rear surface.  
The maximum possible loss compensation strongly depends on the misalignment of the etalon 
rear surface. Already for a misalignment of 1\,urad the compensation range is reduced by
 about 10\,\% and roughly halfed for a misalignment of 3.5\,urad.

\begin{figure}[htb]
\begin{center}
\IG [scale=0.4] {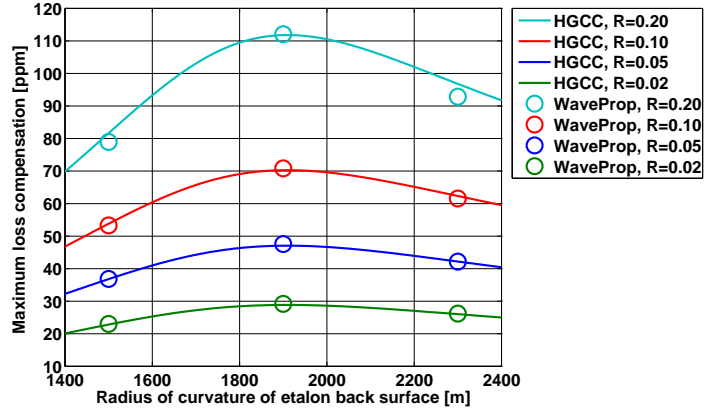} \\
\end{center}
\caption[curvature mismatch]
{The plot shows the maximum differential loss that can be compensated for by the etalon as
a function of the radius of curvature of the spherical back surface of the etalon. The solid traces
 are computed from Hermite-Gauss coupling coefficients (HGCC), while the data points
 represented by a circle are derived from WaveProp simulations. The results from both
 method show that the system is uncritical with respect to moderate deviations in curvature.}
\label{fig:ETR3}
\end{figure}

Figure~\ref{fig:ETR3} shows the maximum loss that can be compensated for as a function
 of the curvature of the etalon back surface. At $R_C=1910$ the curvature
is matched to the beam and values for compensation are maximised. However, 
even with moderate deviations from a matched curvature the range of the etalon 
is reduced only very little. A deviation of the rear surface radius of curvature of
 about 100\,m reduces the maximum compensation range by only a few percent.

In summary, the tuning range of the etalon and therefore also the maximal achievable loss
compensation range strongly depends on the misalignment (non-parallelism) of the etalon 
rear surface, while the radius of curvature of the rear surface is rather uncritical.
Therefore it will be necessary to set tight parallelism requirements for the polishing 
accuracy of the etalon substrate.   

In case it turns out that with state-of-the-art technology it is impossible to achieve a 
misalignment of the rear surface of less than 1\,urad, one might consider to first measure
the non-parallelism of the polished etalon substrate and afterwards decide on the
actual reflectance of the rear coating, necessary to achieve the
actual  required loss compensation range.

\subsection{Simulation: Influence of etalon effect onto alignment signals}
\label{sec:align}

Imperfections of the etalon do not only decrease the achievable loss compensation, but may 
also harm other detector subsystems due to the additional creation of higher order optical
modes. Since the strongest disturbances are expected to show up in the global alignment
 sensing and control system of Advanced Virgo, we performed some detailed \textsc{Finesse}
 simulations
of the alignment control signals using an imperfect etalon. 

 The effect of a possible mismatch between the
alignment of the back and front face of the etalon was found to have very small influence
onto the alignment error signals of the arm cavity.  The coupling of etalon rear surfaces
 misalignment into the arm cavity alignment signals is about 4 to 5 orders of magnitude
 below the coupling of the etalon front surface misalignment \cite{Virgo-note-alignment}.
Also the amount of first order modes (TEM$_{01}$ and TEM$_{10}$) originating from
etalon imperfections is found to be negligible \cite{Virgo-note-alignment}.
 Thus from the point of view of the alignment control system there are no evident 
disadvantages of using an etalon as end mirror of the Advanced Virgo  arm cavities.


\section{Summary and Conclusion} \label{sec:summary}
We have presented a new arm cavity configuration for advanced gravitational wave detectors.
Our concept combines the advantages of the different currently used arm cavity designs
and offers maximized flexibility, by providing access to all important interferometer ports
and simultaneously allowing in-situ balancing of the two arm cavities. Detailed calculations 
and numerical simulations have been performed to evaluate the performance of the 
proposed arm cavity design. These analyzes considered ideal optical elements  as well as 
the case of optics with significant imperfections. Our concept is found to have only negligible
implications for the other detector subsystems. Due to the flexibility offered by arm 
cavities featuring the possibility of in-situ tuning of the losses, the authors believe this concept
 might speed up commissioning of the advanced gravitational wave detectors.

\section{Acknowledgment} 

We would like to thank 
Harald L\"uck, Michele Punturo, Giovanni Losurdo and Raffaele Flaminio for fruitful discussions.  This work has been
supported by the Science and Technology Facilities Council (STFC), Max-Planck Society (MPG) and the European Gravitational Observatory (EGO).


\section*{References} \begin{thebibliography}{3}

\bibitem{geo} Hild S \emph{et al} 2006 The Status of GEO\,600 \emph{ Class.
Quantum Grav.} \textbf{23} S643--S651

\bibitem{virgo} Acernese F \emph{et al} 2004 Status of VIRGO \emph{ Class.
Quantum Grav.} \textbf{ 21} S385--94

\bibitem{ligo} Sigg D \emph{et al} 2004 Commissioning of LIGO detectors \emph{
Class. Quantum Grav.} \textbf{ 21} S409--15

\bibitem{tama} Takahashi R (the TAMA Collaboration) 2004 Status of TAMA300
\emph{ Class. Quantum Grav.} \textbf{ 21} S403--8

\bibitem{advligo} \url{http://www.ligo.caltech.edu/advLIGO/}

\bibitem{advvirgo} \url{http://wwwcascina.virgo.infn.it/advirgo/}


\bibitem{bayer} Bayer-Helms F 1984
Coupling coefficients of an incident wave and the modes of a spherical
optical resonator in the case of mismatching 
Appl. Opt. {\bf 23} 1369--1380.


\bibitem{siegman}
 Siegman A E 1986 Lasers, University Science Books, Mill Valley 
 see also the Errata at \url{http://www-ee.stanford.edu/~siegman/lasers_book_errata.pdf}


\bibitem{Meers88} Meers B~J 1988 Recycling in laser-interferometric
gravitational-wave detectors \emph{\ Phys. Rev. D} \textbf{38}  2317

\bibitem{hrg_Dual} Grote H \emph{et al} 2004 	Dual recycling for GEO 600,
Class. Quantum Grav. \textbf{21} No 5,  S473--S480

\bibitem{40m} Miyakawa 0 \emph{et al} 2006   Measurement of optical response of a detuned resonant sideband extraction gravitational wave detector, Phys. Rev. D \textbf{74}, 022001 

\bibitem{lawrence} Lawrence R \emph{et al}  2002  Adaptive thermal compensation of test masses in advanced LIGO, Class. Quantum Grav. 19 1803-1812


\bibitem{Somiya06} Somiya K and Chen Y et al  "Frequency noise and intensity noise of next-generation gravitational-wave detectors with RF/DC readout schemes" \emph{Phys. Rev. D} \textbf{73} (2006) 122005

\bibitem{Somiya07b} Somiya K and Chen Y et al  "Erratum: Frequency noise and intensity noise of next-generation gravitational-wave detectors with RF/DC readout schemes" \emph{Phys. Rev. D} \textbf{75} (2007) 049905


\bibitem{Braginski04}  
{Braginski} V.~B and {Vyatchanin} S.~P  "Corner reflectors and 
quantum-non-demolition measurements in gravitational wave antennae" 
\emph{Physics Letters A} \textbf{324} (2004) 345-360

\bibitem{Virgo-note-phase-noise} Hild S and Freise A "Advanced Virgo design: Optical phase
 noise originating from the etalon effect driven by thermo-refractive noise" Virgo note
 VIR-058A-08, available at {\tt https://pub3.ego-gw.it/codifier/}

\bibitem{Waveprop} Available at \url{http://www.rzg.mpg.de/~ros/WaveProp/}

\bibitem{Virgo-note-alignment} Mantovani M \emph{et al} "The consequences of using the
 etalon effect to tune the arm cavity finesse on the alignment signals for Advanced Virgo" 
Virgo note VIR-027A-08, available at {\tt https://pub3.ego-gw.it/codifier/}  

\bibitem{fft}
J.-Y.~Vinet, P.~Hello, C.N.~Man and A.~Brillet
J.~Phys.~I (Paris) \textbf{2} (1992) 1287

\bibitem{finesse1}
Interferometer simulation software \textsc{Finesse}, available at \url{http://www.rzg.mpg.de/~adf/}

\bibitem{finesse2}
Freise A  \emph{et al}, 2004, Frequency-domain interferometer
simulation with higher-order spatial modes, \emph{ Class. Quantum
Grav.} \textbf{ 21} S1067-1074.

	

\end{thebibliography}
\end{document}